\documentclass[a4paper,preprint,unsortedaddress,onecolumn]{revtex4-1}
\usepackage{geometry} 
\usepackage{enumerate}
\usepackage{amsmath}
\usepackage{amssymb}
\usepackage[dvips]{graphicx}
\usepackage{color}
\usepackage{tabularx}
\usepackage{gensymb}
\usepackage{multirow}

\usepackage[footnotesize]{subfigure}
\newcommand{\redc}[1]{{\color{black} #1}}
 
\newcommand{\vect}[1]{\textbf{\textit{#1}}}

\usepackage{color}
\definecolor{light}{gray}{0.50}
\definecolor{heavy}{gray}{0.35}
\definecolor{black}{gray}{0.0}
\definecolor{dgreen}{rgb}{0.0,0.7,0}
\definecolor{dred}{rgb}{0.9959,0,0}
\definecolor{green}{rgb}{0.0,0.99599,0.0}
\definecolor{purple}{rgb}{0.6,0.0,0.4}

\newcommand{\blue}[1]{\textcolor{black}{#1}}

\usepackage{smaller}

\newcounter{rupertcommentno}
\newcounter{luigicommentno}
\newcommand{\dss}{\displaystyle}
\newcommand{\bigoh}[1]{\mathcal{O}\left(#1\right)}

\begin{document}
\title{Adaptive Molecular Resolution Approach in Hamiltonian Form: An Asymptotic Analysis}
\author{Jinglong Zhu}
\affiliation{Institute for Mathematics, Freie Universit\"at Berlin, Germany\\
School of Mathematical Sciences, Peking University, Beijing, 100871 P.R.China}
\author{Rupert Klein}
\affiliation{Institute for Mathematics, Freie Universit\"at Berlin, Germany}
\author{Luigi Delle Site}
\affiliation{Institute for Mathematics, Freie Universit\"at Berlin, Germany}
\email{luigi.dellesite@fu-berlin.de}
\begin{abstract}
Adaptive Molecular Resolution approaches in Molecular Dynamics are becoming relevant tools for the analysis of molecular liquids characterized by the interplay of different physical scales. The essential difference among these methods is in the way the change of molecular resolution is made in a buffer/transition region. In particular a central question concerns the possibility of the existence of a global Hamiltonian which, by describing the change of resolution, is at the same time physically consistent, mathematically well defined and numerically accurate. In this paper we present an asymptotic analysis of the adaptive process complemented by numerical results and show that under certain mathematical conditions a Hamiltonian, which is physically consistent and numerically accurate, may exist. \blue{Such conditions show that molecular simulations in the current computational implementation require systems of large size and thus a Hamiltonian approach as the one proposed, at this stage, would not be practical from the numerical point of view. However, the Hamiltonian proposed provides the basis for a simplification and generalization of the numerical implementation of adaptive resolution algorithms to other molecular dynamics codes.}
  
\end{abstract}

\maketitle


\section{introduction}
Large systems of molecular liquids are characterized by processes occurring at different scales which in turn often require a different level of accuracy regarding the molecular model \cite{entropy}. Highly accurate molecular models lead to a complete physical picture but require large computational resources and additional work of analysis of the large amount of data produced; in fact a clear description of a process requires a screening of data to the essential. On the other hand, less accurate molecular models are computationally convenient and produce a small amount of data to analyze, however they are likely to miss, due to their simplification, essential physical features \cite{annurev}.
In this perspective, multiscale methods in Molecular Simulation (MS) have been developed in the last years for optimizing the need of a consistent physical treatment and acceptable numerical and analysis costs \cite{entropy,annurev}. In particular adaptive molecular resolution methods, which by partitioning the system in regions of different molecular resolution, change the number of degrees of freedom on-the-fly became very appealing due to successful numerical performance \cite{jcp1,ensing1}. \blue{Various implementation of this method exist and they differ for the definition of the coupling between different regions \cite{jcp1,ensing1,hadress1,njp}. In this work we propose a procedure for coupling the atomistic and coarse-graining region via an interface region which acts as a filter to transform atomistic resolution into coarse-graining resolution and vice versa. The coupling is done by considering the Hamiltonian of the interface as a perturbation to an otherwise exact Hamiltonian written as a sum of a full atomistic and a full coarse-grained Hamiltonian. Next we perform a mathematical treatment in terms of asymptotic analysis involving characteristic lengths of the system. In this way we derive mathematical conditions which, if reasonably fulfilled in the numerical simulation, assure that the system behaves practically as a 
Hamiltonian system. The existence of a global Hamiltonian, although not necessary for adaptive resolution simulations \cite{prx,njp}, could provide technical advantages in the implementation of the code. In fact all standard molecular dynamics codes are based on Hamiltonian algorithms and thus one may use their computational architecture in an almost straightforward way. In general the direct coupling of an atomistic system with a coarse-graining system is expected to lead to dissipation thus different correction terms are added to take care of such a problem \cite{jcpsimon,ensing1,prlres,hadress1,prx,njp}. In particular in Hamiltonian-based algorithms the Hamiltonian is corrected by adding a free-energy term. Our aim instead is to provide a definition of global Hamiltonian based solely on particles' degrees of freedom, it is our opinion that a formula that contains free-energy terms does not define a proper Hamiltonian. If such a definition is possible, then the system would be self-contained in the sense that once the simulation set up is defined, the numerical calculation can run without any need of further additional quantity calculated in additional simulations. In particular our partitioning of the system paves the way for a generalization of the algorithm to different codes. This can be done by a simplification of the computational algorithm offered by the partitioning we propose: the simulation (given the interaction cutoff) at a certain stage can be performed in two distinct regions (atomistic and coarse-grained) and then a third region where they overlap (the region of perturbation). For each time step, force calculations (i.e. the most expensive part of the code) can effectively run in parallel in each of the two regions and then synchronized through the force calculation in the overlapping region. Work along this direction is currently in progress. Finally, it must be also clarified that all the adaptive methods cited before \cite{ensing1,hadress1,prx,njp} from the mere numerical point of view are essentially equivalent and/or equivalently efficient. The conceptual difference we discuss here regards the formal background on which they are based and the physical interpretation of global Hamiltonian in terms of statistical mechanics of their results (see also note 4 in Ref.\cite{njp})}


\section{Adaptive Molecular Resolution: Force-based or \\ Hamiltonian-based}
In this section we review the basic principles employed in the construction of adaptive resolution schemes. The schemes can be classified in two major categories: (a) Force-based interpolation schemes \cite{jcp1,annurev}; (b) Potential-based (Hamiltonian-based) schemes \cite{ensing1,ensing2,ensing3,bulo1,hadress1,eplhadress}. 


\subsection{Force-based and Grand-Canonical-like scheme}
\begin{figure}[htbp]
\begin{center}
\includegraphics[width=7.5cm]{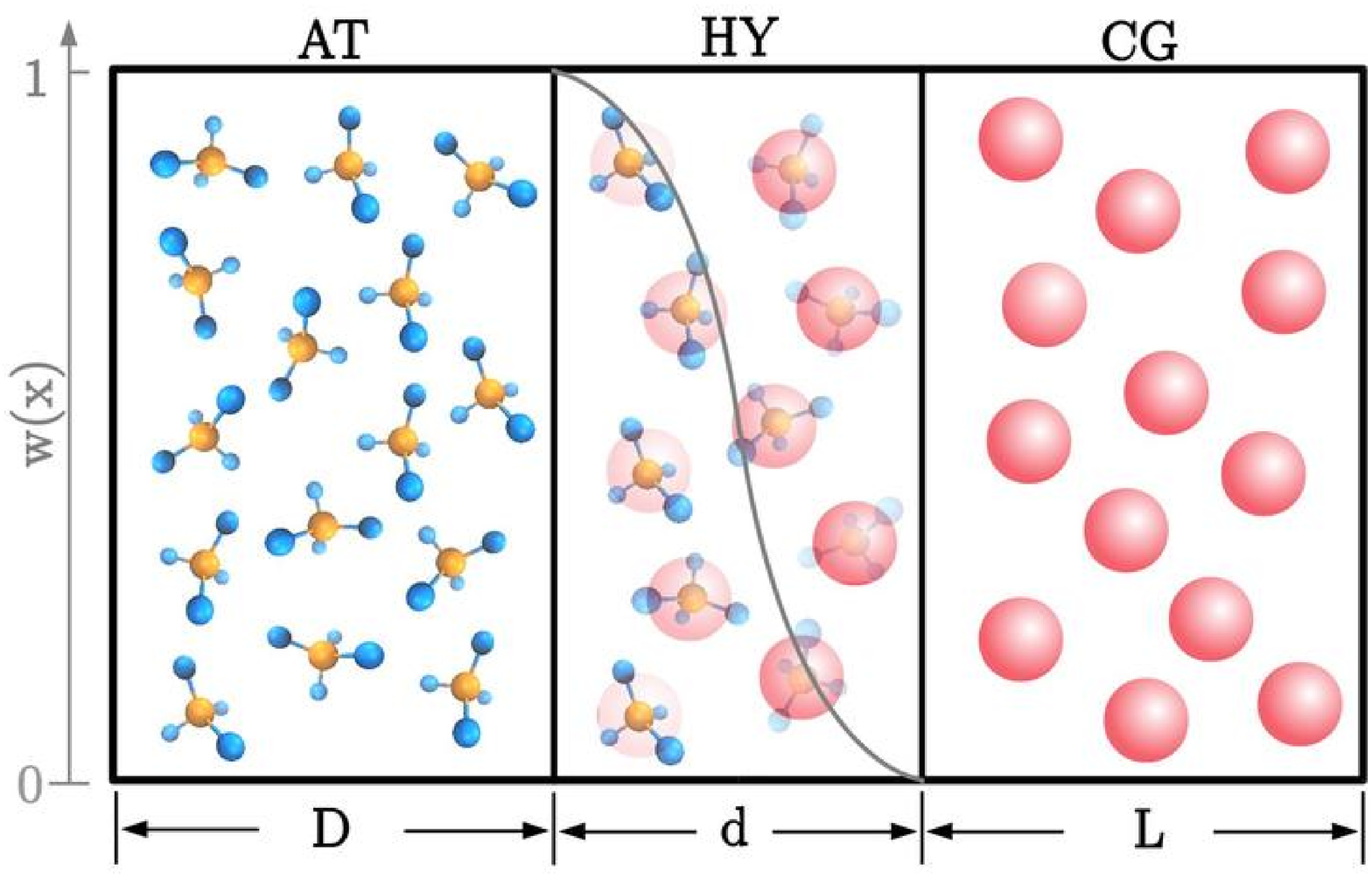}
\caption{Pictorial representation of the adaptive resolution set up.} \label{system-setup}
\end{center}
\end{figure}

The Adaptive Resolution Simulation (AdResS) method has been developed following a simple intuitive principle. Such a principle consists of dividing the space in three distinct regions: (i) atomistic region (high resolution), (ii) coarse-grained region (low resolution), (iii) interface or hybrid/transition region where molecules change their resolution. Next the intuitive physical requirement is that the molecules of the atomistic region follow the rules of a standard atomistic dynamics, the  molecules of the coarse-grained region  follow the rules of a standard coarse-grained dynamics and when a molecule transits in the interface region its dynamics slowly passes from atomistic type to a coarse-grained type (or vice versa), The meaning of ``slowly'' is that the perturbation due to the change of resolution to the dynamics of the atomistic and of the coarse-grained region is negligible in the calculations of physical quantities of interest. In Molecular Dynamics (MD) such a principle can be 
easily implemented by smoothly interpolating in space the atomistic and coarse-grained forces:
\begin{equation}
{\vect F}_{i,j}=w(\vect r_i)w(\vect r_j){\vect
  F}_{i,j}^{AT}+[1-w(\vect r_i)w(\vect r_j)]{\vect F}^{CG}_{i,j} 
\label{eqforce}
\end{equation}
where $i$ and $j$ are the indices of two distinct molecules, ${\vect F}^{AT}$ is the force derived from the atomistic potential ($U_{AT}$) and  ${\vect F}^{CG}$
is the force derived from the coarse-grained potential ($U_{CG}$) (usually a COM-COM potential, where COM indicates ``the center of mass''), $\vect r$ is the COM position of the molecule and $w(x)$ is a smooth function varying from $0$ to $1$ in the transition region ($HY$):
\begin{equation}
    w(x) = \begin{cases}
               1               & x < x_{AT} \\
               cos^{2}\left[\frac{\pi}{2(d)}(x-x_{AT})\right]   & x_{AT} < x < x_{AT}+d\\
               0 & x_{AT} + d< x
           \end{cases}
\end{equation} 
with $x_{AT}$ the x-location of the border of the AT region (see Fig.\ref{system-setup}). 
In an effective way, the atomistic degrees of freedom are slowly removed when a molecule leaves the atomistic region and enters into the coarse-grained region and vice versa.
As it is described above, the scheme is dissipative, that is the change in number of degrees of freedom implies a gain or loss of kinetic and potential energy which is not spontaneously balanced, and in fact the system, prepared following this scheme, is not Hamiltonian \cite{prlcomm}. However, when coupled to a an external thermostat, which takes care of adsorbing the excess or adding the missing energy, it was proven to be technically sufficient for the scheme to work properly \cite{prefract,jpaadress}. At a later stage this empirical finding was justified on the basis of first principles of thermodynamics and statistical mechanics and recently the basic scheme reported above was embedded into the formalism of a Grand Canonical ensemble where the coarse-grained and transition region act as a particle reservoir for the atomistic region \cite{jcpsimon,prlres,jctchan,prx,njp,jcppi,pre2}. The technical details are not relevant for this paper, except that in a Grand Canonical-like set up, the reservoir acts in 
a stochastic way and thus the only Hamiltonian of relevance is that of the atomistic region which implies that the non-existence of a global Hamiltonian is no more a problem. However, some branches of research followed a different path based on the search of a global Hamiltonian without the addition of any stochastic aspect. An overview of the progress and pitfalls of such approaches is reported below. 


\subsection{Hamiltonian based schemes}
In a series of papers Ensing and collaborators \cite{ensing1,ensing2,ensing3,bulo1} presented a scheme based on the interpolation of potentials instead of forces as done in AdResS:
\begin{equation}
U_{ij}^{tot}=s(\vect r_i,\vect r_j)
  U_{i,j}^{AT}+[1-s(\vect r_i,\vect r_j)]U^{CG}_{i,j}
\label{ham}
\end{equation}
where $U_{ij}$ is the global potential of interaction between molecule $i$ and molecule $j$, $s(\vect r_{i}, \vect r_{j})$ is the smooth interpolation function (slightly but not substantially different from the function $w$ used in AdResS) and $ U_{i,j}^{AT}$, $U^{CG}_{i,j}$ are the atomistic and coarse-grained potential acting between molecule $i$ and molecule $j$. If, from $U_{ij}^{tot}$ one derives the force between molecule $i$ and molecule $j$ then the result is a force as that of AdResS plus an additional term: $\nabla s(\vect r_{i}, \vect r_{j})(U_{i,j}^{AT}-U^{CG}_{i,j})$. In their work \blue {the dissipative action of this spurious force needs to be balanced such that it disappears from the simulation statistics. To achieve this, the method uses book-keeping of the gain/loss of energy} of molecules that change resolution and thus adding at each instant the gain/loss of energy due to the change of resolution and preserve a global Hamiltonian approach.
\blue{The accurate determination of the gain/loss of energy per molecule and its adsorption/release in the system was done by violating the request that the system should be self-contained.} In fact, in essence, the system was coupled to a generic  external thermostat and thus this method became equivalent to the force-based AdResS scheme. It was then shown that without the thermostat the system was highly dissipative and that the interpolation of potentials plus the book-keeping energy was not sufficient to build a conservative scheme \cite{prlcomm}. However, later on, the idea of providing/removing energy without using a thermostat was technically implemented by a method named ``H-AdResS'' developed by Potestio and collaborators \cite{hadress1,eplhadress,jcphadress,h-addressmc}. In H-AdResS the interpolation is also done as in Eq.\ref{ham} and it was observed that such an interpolation is equivalent to the scheme of 
thermodynamic integration for the calculation of the difference of free energy between a state A and a state B of the system. In this case state A and state B meant the passage from the atomistic potential to the coarse-grained potential and thus the interpolation formula is equivalent to the space dependent change of free energy between the atomistic and the coarse-grained representation:
\begin{equation}
\Delta F(\lambda)=\int_{0}^{\lambda}<U^{AT}-U^{CG}>_{\lambda^{'}}d\lambda^{'}
\label{h-ad}
\end{equation}
where $<...>$ indicates the ensemble average and each $\lambda$ corresponds to fixed values of the switching function $w(x)$ in the transition region, thus $\Delta F(\lambda)=\Delta F(\lambda(x))$.
Thus the Hamiltonian was {\it a posteriori} modified so that the balancing term of free energy, $\Delta F(\lambda)$ was added in an effective Hamiltonian. This implies that there is no need of a generic and undefined book-keeping of the molecular energy, but the missing energy can be quantified by $\Delta F(\lambda)$. However \blue{there are several conceptual problems with such an approach}. There are at least three arguments for this thesis: (i) The scheme violates third Newton's law, in fact the force term:  $\nabla w(\vect r_{i})w( \vect r_{j})(U_{i,j}^{AT}-U^{CG}_{i,j})$ is not antisymmetric in $i$ and $j$ \cite{jpaadress}; (ii) The Hamiltonian is ill defined because any additional/corrective one particle potential, like $\Delta F(\lambda(x))$, must be a solution of a partial 
differential equation of the first order but with two boundary conditions \cite{pre1}; (iii) the effective Hamiltonian is not a first principles Hamiltonian but, by construction, depends on the specific thermodynamic state point (in fact it carries a free energy term: $\Delta F(\lambda(x))$) \cite{prx,entropy,njp}. In this perspective the search for a first principle (self-contained) Hamiltonian that allows for a change of resolution without the need of {\it a priori} or {\it a posteriori} (artificial) corrections, motivated us to explore new directions, as reported in the next section.


\section{Reformulation of the Transition region via Asymptotic Analysis}


\subsection{Symbols and labels}
Table.\ref{tab3} is a list of symbols and labels used in this section for future reference:
\begin{table}[htbp] 
\caption{\blue{Symbols and labels}}
           \centering
           \begin{tabular}{|c|l|} 
           \hline 
 	   Symbol &  Meaning\\
           \hline
           $V_{AT}$ & Atomistic potential\\
           \hline
           $V_{CG}$ & Coarse-grained potential \\
           \hline
           $F_{AT}$ & Atomistic force \\
           \hline
           $F_{HY}$ & Hybrid force \\
           \hline
           $F_{CG}$ &  Coarse-grained force\\
           \hline
           a, b  &  label of molecules (a, b=1, 2, 3,$\cdots$, $N_{tot}$) \\
           \hline
           $N_{tot}$ & Total number of molecules  \\
           \hline
           ia & i-th atom of the a-th molecule \\
           \hline
           sa &  number of atoms in a-th molecule\\
           \hline
           ${\bf r}_{a}$ &  set of spatial coordinates of all atoms composing the a-th molecule\\
           \hline
           $m_{ia}$ & mass of the i-th atoms of the a-th molecule \\
           \hline
           ${\bf R}_{a}$ &$(X_{a},Y_{a},Z_{a})$, coordinates of the center of mass of the a-th molecule  \\
           \hline
           ${\bf p}_{ia}$ & Momentum of the i-th atom of the a-th molecule \\
           \hline
           $N_{AT}$ & number of molecules in AT region \\
           \hline
           $N_{CG}$ &number of molecules in CG region  \\
           \hline
           $N_{HY}$ & number of molecules in HY region  \\
           \hline
           D &  characteristic extent of AT region along the change of resolution\\
           \hline
           d & characteristic extent of HY region along the change of resolution \\
           \hline
           L & characteristic extent of CG region along the change of resolution \\
           \hline
           \end{tabular}
           \label{tab3}
\end{table}

\subsection{Perturbed Hamiltonian via asymptotic expansion}
Consider a typical adaptive resolution set up as illustrated in 
Fig.\ref{system-setup} and let us decompose the hypothetical global Hamiltonian of the system as:
\begin{equation}
H_{glob}=H_{AT-AT}+H_{CG-CG}+H_{HY-HY}+H_{AT-HY}+H_{CG-HY} + H_{intra}
\label{hglob}
\end{equation}
where:
\begin{equation}
H_{AT-AT}=\sum_{a=1}^{N_{AT}}\left(\sum_{ia=1}^{sa}\frac{{\bf p}^{2}_{ia}}{2m_{ia}}\right)+\sum_{a=1}^{N_{AT}}\left(\sum_{b=1; b\neq a}^{N_{AT}}V_{AT}({\bf r}_{a},{\bf r}_{b})\right)
\label{hatat}
\end{equation}
\begin{equation}
H_{CG-CG}=\sum_{a=1}^{N_{CG}}\left(\sum_{ia=1}^{sa}\frac{{\bf p}^{2}_{ia}}{2m_{ia}}\right)+\sum_{a=1}^{N_{CG}}\left(\sum_{b=1; b\neq a}^{N_{AT}}V_{CG}({\bf R}_{a},{\bf R}_{b})\right)
\label{hcgcg}
\end{equation}
\begin{eqnarray}
H_{HY-HY}&=&\sum_{a=1}^{N_{HY}}\left(\sum_{ia=1}^{sa}\frac{{\bf p}^{2}_{ia}}{2m_{ia}}\right)+\left[\sum_{a=1}^{N_{HY}}\left(\sum_{b=1; b\neq a}^{N_{HY}}w(X_{a})w(X_{b})V_{AT}({\bf r}_{a},{\bf r}_{b})\right)\right]\nonumber\\ &+&\left[\sum_{a=1}^{N_{HY}}\left(\sum_{b=1; b\neq a}^{N_{HY}}(1-w(X_{a})w(X_{b}))V_{CG}({\bf R}_{a},{\bf R}_{b})\right)\right]
\label{hhyhy}
\end{eqnarray}
\begin{equation}
H_{AT-HY}=\sum_{a=1}^{N_{AT}}\left(\sum_{b=1; b\in HY}^{N_{HY}}V_{AT}({\bf r}_{a},{\bf r}_{b})\right)
\label{hathy}
\end{equation}
\begin{equation}
H_{CG-HY}=\sum_{a=1}^{N_{CG}}\left(\sum_{b=1; b\in HY}^{N_{HY}}V_{CG}({\bf R}_{a},{\bf R}_{b})\right).
\label{hcghy}
\end{equation}
\begin{equation}
H_{intra}=H_{bond}+H_{angle}
\end{equation}
Where $H_{bond}$ is the sum of all intramolecular atom-atom bonding energies, and $H_{angle}$ is the sum of all the intramolecular bonding angle energies. The specific form of $H_{intra}$ depends on the molecular model used and it will be defined in Section IV.

In the formulas above we have worked from the hypothesis that the atomistic degrees of freedom for the kinetic energy and of the intramolecular potentials are present in all molecules independently of the resolution \blue{(double resolution everywhere)} but their adaptive character is considered only in relation to the intermolecular interaction sites. Such a situation corresponds to the actual (current) numerical implementation of the adaptive resolution scheme, in fact from the numerical point of view the calculations of the forces correspond to about $75\%$ of the computational effort. Moreover, the coarse-grained potential is derived to reproduce the  thermodynamics and the probability distribution function in space up to the two body case, i.e., the radial distribution function $g({\bf r})$, of the atomistic resolution.
At this point we can rewrite Eq.\ref{hglob} as the sum of ``exact'' and ``perturbation'' terms as follows. The exact terms involve all contributions to the energies and interactions in Eq.\ref{exact} that are well defined by the physics and do not involve any external or artificial quantities, such as $w(x)$. Thus,
\begin{equation}
H_{exact}=\left[H_{AT-AT}+H_{AT-HY}\right]+\left[H_{CG-CG}+H_{CG-HY}\right]+H_{intra}+K_{HY}\,,
\label{exact}
\end{equation}
where $K_{HY}=\sum_{a=1}^{N_{HY}}\left(\sum_{ia=1}^{sa}\frac{{\bf p}^{2}_{ia}}{2m_{ia}}\right)$ is the kinetic energy of the hybrid region. The remaining perturbations to the exact Hamiltonian related to the presence of the hybrid region read:
\begin{eqnarray}
\dss \Delta H
  & =
    & \dss 
      \left[\sum_{a=1}^{N_{HY}}
        \left(\sum_{b=1; b\neq a}^{N_{HY}} w(X_{a})w(X_{b})V_{AT}({\bf r}_{a},{\bf r}_{b})
        \right)
      \right]
      \\[20pt]
  & +
    & \dss 
      \left[\sum_{a=1}^{N_{HY}}
        \left(\sum_{b=1; b\neq a}^{N_{HY}}(1-w(X_{a})w(X_{b}))V_{CG}({\bf R}_{a},{\bf R}_{b})
        \right)
      \right]. \label{pert}
\end{eqnarray}

Let us now reformulate the definition of the switching function $w(x)$ \redc{in} terms of some characteristic lengths of the problem:
\begin{equation}
w(X)=\hat{w}\left(\phi(X)/d\right)=\hat{w}(\xi)
\label{neww}
\end{equation}
where $\phi(X)$ is the signed distance from the boundary of the atomistic region and $d$ is the characteristic thickness of the hybrid/transition region, \emph{i.e.}, the hybrid region is covered letting $0\leq \xi \leq 1$. One of the characteristic lengths of primary importance is $l_{c}$, that is, the range of molecular interaction. Other characteristic lengths are the size of the AT region, $D$, and that of the CG region, $L$. 

To proceed in the development of an asymptotic limit formulation, and to meet the actual numerical set up used in AdResS simulations, we assume to work in a regime with $\epsilon=l_{c}/d<<1$. 
We observe that the atomistic and coarse-grained interatomic forces, $F_{ia,jb}^{AT}$ and $F_{ab}^{CG}$, are very small when $|X_{a}-X_{b}|> l_{c}$. As a consequence, the weighting functions $w(X_{a}), w(X_{b})$ in the expression for forces in the hybrid region can only contribute sizeably when $|X_{a} - X_{b}| \lesssim l_c$. Thus, whenever the intermolecular forces are sizeable, a Taylor expansion of the weighting function evaluated at $\bar X=\frac{X_{a}+X_{b}}{2}$ yields, e.g., for $X_a$,
\begin{equation}
w (X_{a}) = w(\bar X) + \epsilon \frac{d \hat w}{d \xi} \frac{X_a -\bar X}{l_c}  \nabla \phi (\bar X) + \bigoh{\epsilon^2}
\label{wo}
\end{equation}
and the perturbation Hamiltonian reads
\begin{equation}
\Delta H = \Delta H(\epsilon) = w(\bar X)^2 V_{AT} + (1-w(\bar X)^2) V_{CG} + \bigoh{\epsilon}\,.
\label{ho}
\end{equation}
Since $w(\bar X)$ is slowly varying, i.e., $\frac{dw(x)}{dx} = \bigoh{\epsilon}$, we obtain the 
leading force term, 
\begin{equation}
\vec F_{HY} = w(\bar X)^2 \vec F_{AT} + (1-w(\bar X)^2) \vec F_{CG}+\bigoh{\epsilon}\,.
\label{newf}
\end{equation}
%
This is a force formulation for the transition region, that is close to that known from the standard adaptive resolution approaches discussed before which was never derived before from a potential energy/Hamiltonian point of view. Here, this force emerges as the gradient of the interpolated potential up to perturbations of order {$\epsilon$}. As a consequence, the particle motion within the transition region follows some perturbed weakly non-Hamiltonian dynamics. 
This is the starting point of a further asymptotic analysis regarding different scales involved in the problem, in particular, it becomes of interest to perform such an analysis with respect to the other characteristic lengths involved, such as $D$ and $L$. $D$, the size of the AT region, represents in principle the smaller scale, that is, it is the region involved in the observation of a very local event, looked at with all atomistic details; instead $L$, the extension of the coarse-grained region, represents the scale of the part of the system with the role of a large macroscopic reservoir which ensures that the macroscopic quantities of the thermodynamics state (particle density, temperature, pressure) are preserved. In the following we explain how to use Eq.\ref{newf} and the asymptotic approach for $L$, $d$ and $D$ to determine the degree of the perturbation and thus, if this is negligible, to identify the conditions under which the adaptive scheme is essentially Hamiltonian.


\subsection{Hamiltonian versus Dissipative scheme}
Let us define the quantities $\eta=\frac{D}{L}; \zeta=\frac{d}{L}; \lambda=\frac{D}{d}$. We can the perform numerical simulations using Eq.\ref{newf} and sample the space of $\eta,\zeta,\lambda$ to check numerically combinations of such parameters for which the system is dissipative or Hamiltonian. The ideal set up, from the mathematical point of view, for a Hamiltonian-like behaviour, would be: $\eta<<1;\zeta<<1; \lambda<<1$. In fact one could have a large coarse-grained region which, being dominant assures that the overall thermodynamic conditions are preserved, and a large HY region compared to the AT region where the condition $\frac{dw(x)}{dx} \ll 1$ can be reasonably met. Moreover, such a situation is also optimal for the physical interpretation of multiscale analysis, that is a very localized event analyzed in a very small (AT) region compared to the rest of the system. 
However, since the asymptotic considerations described above rely predominantly on the smallness of $\epsilon = l_c/d$, \emph{i.e.}, on the hybrid region being ``thick'' in comparison with the molecular interaction distance, the case of a large $D$ and a small $d$ would be equally acceptable from this point of view as long as $\epsilon \ll 1$ is guaranteed. In this regime, the large coarse-grained and atomistic regions are expected to produce more reliable spatially nearly homogeneous statistics, although related simulations will, of course, be substantially more expensive.
In the numerical simulation we will have as a reference the full atomistic simulation in a microcanonical ensemble (NVE), that is, the system is self-contained and there \redc{is} no external thermostat (as it is instead in the NVT \redc{)}. We will compare the full atomistic simulation with the adaptive resolution simulation; this latter will also be performed without the help of any external tool/thermostat. By monitoring the temperature as a function of time we check whether the system behaves strongly in a dissipative way or closer to conservative, if the system is dissipative then it is certainly \blue{not} Hamiltonian. If the system is conservative we then check whether structural properties of the atomistic region are the same as those of a full atomistic simulation and structural properties of the coarse-grained region are the same as those of a full coarse-grained simulation; if the result is positive then, from a physical point of view, we can claim to have found a Hamiltonian that by allowing a 
spatial adaptive molecular 
resolution can \redc{preserve} basic thermodynamic properties (i.e. temperature conservation) and structural properties of a full atomistic and full coarse-grained simulation.


\section{Numerical Simulation}


\subsection{Molecular Model}

\begin{figure}[htbp]
\begin{center}
\includegraphics[width=4cm]{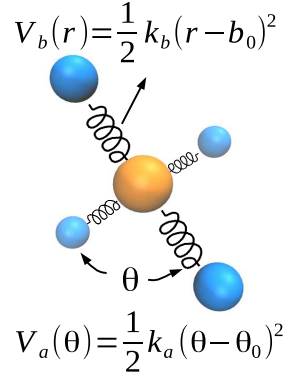}
\caption{Intramolecular Interactions of the full atomistic model} \label{fig2}
\end{center}
\end{figure}
\begin{figure}[htbp]
\begin{center}
\includegraphics[width=4cm]{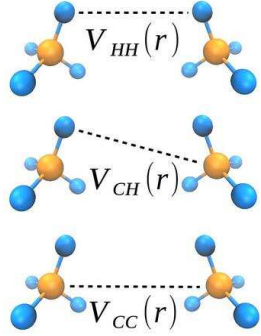}
\caption{Intermolecular Interactions of the full atomistic model} \label{fig3}
\end{center}
\end{figure}
We have constructed a toy model, that is a molecule with a methane-like structure which has a tetrahedral arrangement of lighter atoms (hydrogen-like, thus named H) connected via flexible (harmonic) bonds to a central heavier atom (carbon-like, thus named C) and with the H-C-H bond angle that is also described by a harmonic potential (see Fig.\ref{fig2}). Intermolecular interactions are described by a smoothed Lennard-Jones 12-6 potential{\cite{switch}}:
\begin{equation}
 V_{XY}(r) = C^{XY}_{12} f_{12}(r) - C^{XY}_{6} f_{6}(r)
\end{equation}
while X,Y can be C or H. 
\begin{equation}
    f_s(r) = \begin{cases}
               \frac{1}{r^s} - C_s  &r < r_1\\
               \frac{1}{r^s} - s \frac{A_s}{3} (r-r_1)^3 - s \frac{B_s}{4} (r-r_1)^4 - C_s &r_1 \leq r < r_c\\
               0 &r_c \leq r
           \end{cases}
\end{equation} 
where $s$ can be 12 or 6, $r_1=1.1 $ nm is the switching radius for the smoothing and $r_c=1.3 nm$ is the cut-off radius of interaction. $A_s,B_s$ and $C_s$ are chosen that $f'_s(r_c)=0$ , $f''_s(r_c)=0$ and $f''_s(r)$ is continuous at $r=r_1$. So the original L-J interaction is changed smoothly from $r=r_1$ to zero at $r=r_0$. $C^{XY}_s, k_a, k_b,b_0, \theta_0$ are taken from the Gromacs OPLS/AA force field \redc{\cite{oplsaa-1,oplsaa-2}}. The choice of such a molecular model  is justified by the fact that we need a model sufficiently simple for a straightforward numerical implementation but at the same time also sufficiently complex to represent a valid challenging test for the mathematical principles behind the method. 
With this model, we can now write the specific form of $H_{intra}$ 
\begin{equation}
H_{intra}=\sum_{a=1}^{N_{tot}} \left(\sum_{i=1}^{4}\frac{1}{2}k_b(|\textbf{r}_a^C-\textbf r^{H_i}_a|-b_0)^2+ 
\sum_{i=1}^3 \sum_{j=i+1}^{4}\frac{1}{2}k_a(\theta_a^{H_iCH_j}-\theta_0)^2 \right)
\label{hintra}
\end{equation}
where ${\bf r}_{a}^C$ is the spatial coordinates of the carbon atom of the a-th molecule,
${\bf r}_{a}^{H_i}$ is the spatial coordinates of the i-th hydrogen atom of the a-th molecule,
$\theta_a^{H_iCH_j}$ is the bond angle $H_i$ - C - $H_j$ of the a-th molecule.

We then perform numerical simulations at a temperature 111 Kelvin and pressure 1.0 atm, \emph{i.e.}, a condition at which the structure of this system is typical of a standard liquid.
The coarse-grained model is built by performing the IBI procedure (Iterative Boltzmann Inversion) with pressure correction \cite{ibi}. This procedure assures that the resulting coarse-grained model matches the COM-COM (Center of mass) RDF (Radial distribution function) of the full-atomistic model and assures the same pressure (at the same given temperature). A pictorial representation of the molecular coarse-grained model and the resulting numerical/tabulated potential are shown in Fig.\ref{fig4}.

\blue{
\begin{figure}[htbp]
\begin{center}
\includegraphics[width=12cm]{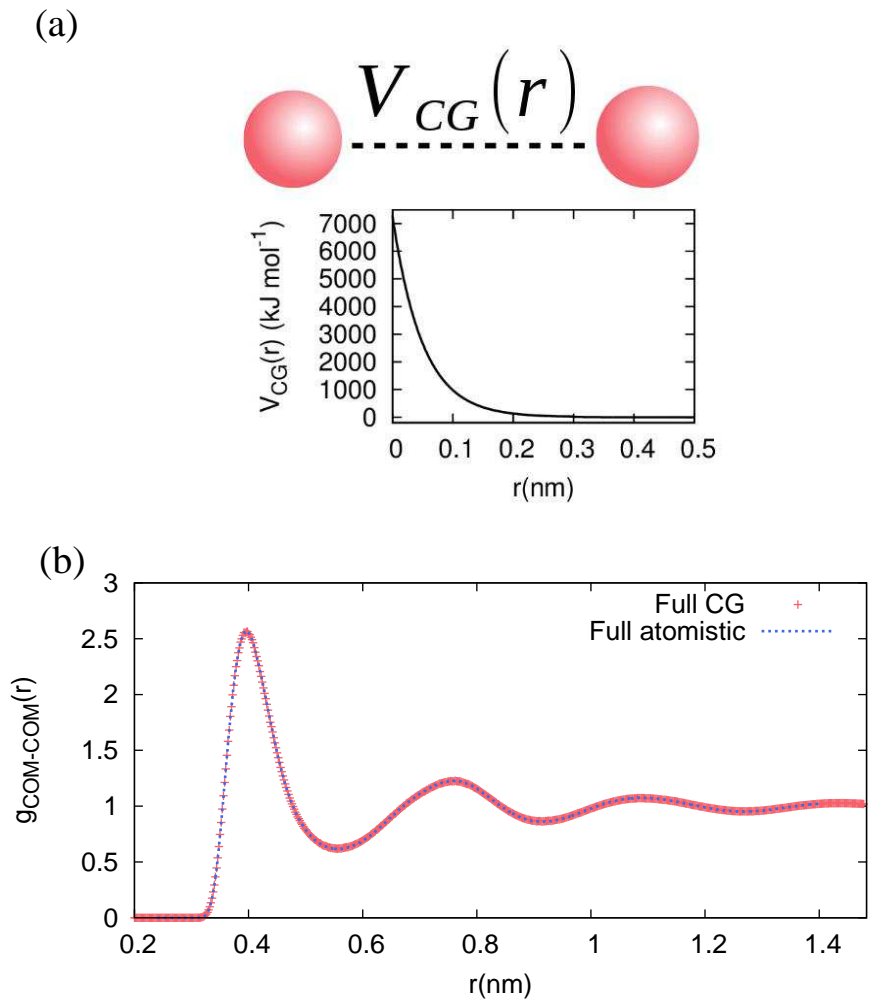}
\caption{\blue{(a) Coarse-grained model and its corresponding numerical potential derived by the IBI procedure. (b) Comparison between the center-of-mass center-of-mass radial distribution function of the full atomistic simulation and the equivalent obtained from the coarse-grained potential of (a) within the IBI procedure.}} \label{fig4}
\end{center}
\end{figure}
}%
At this point we \redc{possess} all the necessary \redc{ingredients} for performing adaptive resolution simulations using Eq.\ref{newf} in the $d$,$D$,$L$ space. We keep $l_{c}$, that is the range of molecular interaction, fixed at a value typical of atomitistic simulations; this choice is taken for practical purposes so that our results can be directly applied to any atomistic simulation of typical molecular liquids of interest in chemical physics. In the next section we report the technical details of \redc{the simulations} so that our results can be reproduced.


\subsection{Technical details}

All simulations are performed using double-precision home-modified Gromacs-4.6.5\cite{gromacs}. The IBI technique with pressure correction is employed to derive the coarse-grained model by using the VOTCA-2.6 package \cite{votca}. 
The time step for the IBI is 0.002 ps and the pressure correction is added every 2 in 3 iteration steps with a standard pressure smoothing at the post update stage. 
The smoothing potential for L-J interaction and for the coarse-grained potential is numerically implemented by using the user defined table with spacing 0.0005 nm. The switching radius of the smoothing is 1.1 nm while the cut-off radius $l_{c}$ is 1.3 nm.


\subsubsection{Preparation/equilibration of the system} 

The system is prepared starting from an initial box containing 470 ``methane'' molecules with dimensions of the system $3.0 \times 3.0 \times 3.0 $ $nm^3$. Next the box is copied along the x-direction to produce the initial configuration for larger systems. 
It follows a full atomistic NPT simulation (that is at fixed number of molecules, pressure and temperature, while the volume is allowed to fluctuate) using the Parrinello-Rahman coupling method \redc{\cite{par-rah-1, par-rah-2}} for 50 ps in order to determine the equilibrium volume. The coupling type is isotropic, the reference pressure is 1.0 atm,  the time constant for pressure coupling is $\tau_{p}= 2.0$ ps and the compressibility is $4.1 \times 10^{-3}$ bar $^{-1}$.
Next, an NVT simulation is thermally equilibrated, with V fixed at the equilibrium value of the NPT simulation, by using a Langevin thermostat for 100 ns. The reference temperature is 110 K and the time constant for temperature coupling is $\tau_{t} = 0.1$ ps. The NVT thermally equilibrated configuration is the used for a full NVE full atomistic simulation (reference calculation) and for an equivalent adaptive resolution simulation. The timestep used for all the simulations is 0.0005 ps.


\subsection{Results and Discussion}

\begin{figure}[htbp]
\begin{center}
\includegraphics[width=6cm]{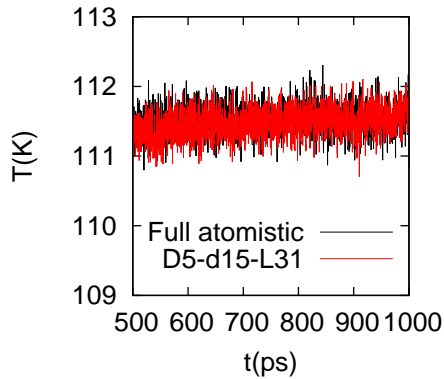}
\caption{Temperature as a function of time for system \blue{(6)}. Data are analyzed after 500 ps to allow the system for basic equilibration in the adaptive set up.} \label{temph2}
\end{center}
\end{figure}

\begin{figure}[htbp]
\begin{center}
\includegraphics[width=6cm]{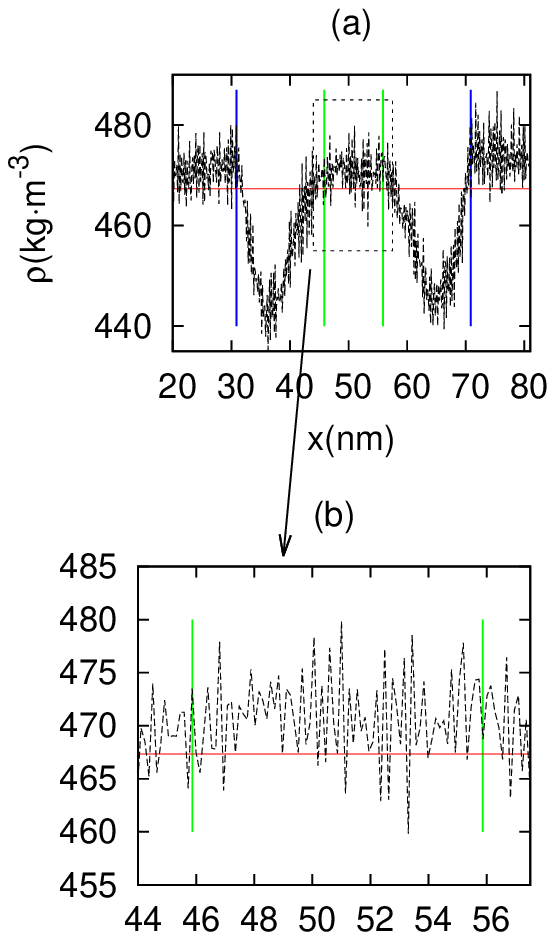}
\caption{\blue{(a)} Molecular number density as a function of the position in the simulation box of system \blue{(6)}. The horizontal line corresponds to the uniform target density. The blue vertical lines correspond to the borders between  the HY and CG region, the vertical green lines correspond instead to to the borders between  the HY and AT region. \blue{(b)} A zoom of the density in the AT region.} \label{rho1}
\end{center}
\end{figure}

\begin{figure}[htbp]
\begin{center}
\includegraphics[width=12cm]{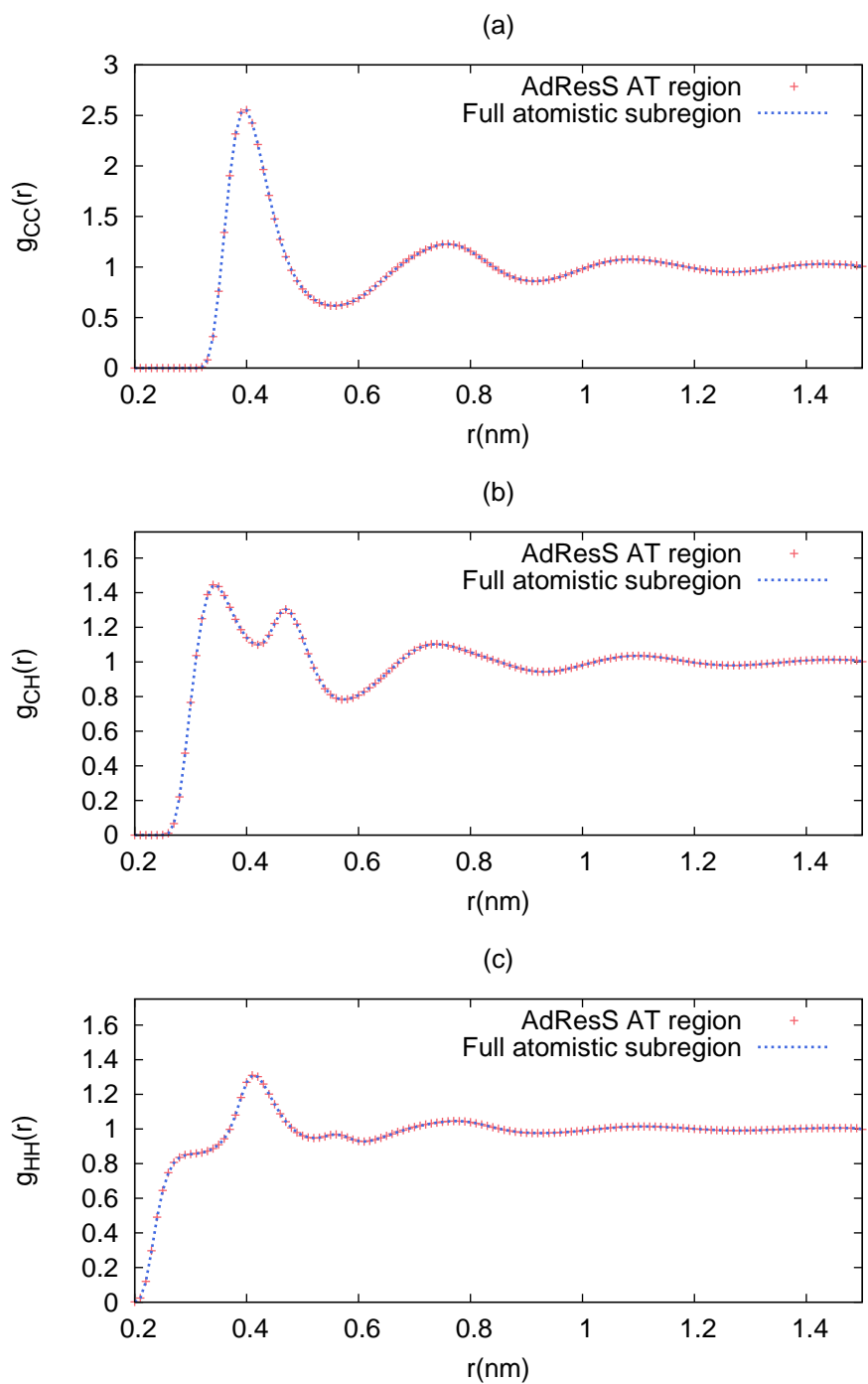}
\caption{Atom-atom radial distribution functions in the AT region compared with the equivalent from the full atomistic simulation of reference.} \label{gofr}
\end{center}
\end{figure}

\begin{table}[htbp] 
\caption{\blue{Systems investigated, $H$ stays for Hamiltonian behaviour while $NH$ for dissipative behaviour. Note that the cut-off radius for intermolecular interaction is always $l_c=1.3nm$.}}
           \centering
           \begin{tabular}{|c|c|c|c|c|c|c|} 
           \hline 
 	   Label & $N_{tot}$ & D(nm) & d(nm) & L(nm) & H or D & $\epsilon$\\
           \hline
           (1) & 47000 & 61.5 &15 & 73.1  & H  & 0.087\\
           \hline
           (2) & 47000 & 13.5 & 3 & 133.5 & NH & 0.433\\
           \hline
           (3) & 28200 & 30  & 30 & 30    & H  & 0.043\\
           \hline           
           (4) & 20680 & 5   & 30 & 31    & H  & 0.043\\
           \hline
           (5) & 18800 & 15  & 15 & 30    & H  & 0.087\\
           \hline
           (6) & 15980 & 5   & 15 & 31    & H  & 0.087\\
           \hline
           (7) & 12220 & 5   & 5  & 29    & NH  & 0.260\\
           \hline
           (8) & 12220 & 5   & 15 & 19    & NH  & 0.087\\
           \hline
           \end{tabular}
           \label{tab2}
\end{table}
Table \ref{tab2} reports results about some representative examples of systems that we have studied by sampling the $D$, $d$, $L$ ($\eta,\zeta,\lambda$) space.
\blue{It must be reported that $\epsilon$ is not the sole parameter governing the $H$ (Hamiltonian) or $NH$ (non Hamiltonian) behaviour. In fact the size of the coarse-grained and of the atomistic region also play an important role. For the atomistic region one would like a size as small as possible (yet statistically valid) so that the saving of the computational resources can be optimized. For the coarse-grained region instead a small size would mean that the distribution of the dissipation of the hybrid region will be distributed among a relatively small number of molecules (that is the coarse-grained region does not act as a large ``reservoir''). As a consequence if the coarse-grained region is too small, the perturbation of the hybrid region will induce a large perturbation (per molecule) of the coarse-grained region and thus a sizable perturbation to the overall thermodynamics of the system.}
Systems of smaller size (i.e. with less than 12000 molecules) are not reported because they do not meet neither the mathematical nor the (intuitive/practical) physical conditions for being Hamiltonian. Instead the systems reported in Table \ref{tab2} are in principle all possible candidates for being Hamiltonian systems, however, some meet the conditions better than others. \blue{The numerical indicator for the classification in $H$ or $NH$ as reported in Table \ref{tab2} is that in the time frame between $0.5$ and $1.0 ns$ the temperature of the adaptive systems overlaps with that of the full atomistic simulation of reference, and that structural properties in the atomistic region (and trivially in the coarse-grained region) agree with those of the corresponding full atomistic simulation of reference. This classification is based on the idea that since after $0.5 ns$ the system is equilibrated, for (at least) short trajectories of $0.5 ns$ (i.e. from $0.5$ to $1.0 ns$) we run in a Hamiltonian regime which fully corresponds to that of full atomistic simulation. Thus for calculating statistical properties one may create an ensemble of independent trajectories of (at least) length $0.5 ns$ and obtain valid results by averaging over the ensemble. In previous adaptive resolution studies such approach (even over shorter time windows) has been already used \cite{cpc}. However, in \ref{beyond} we will also discuss one example where the simulation run is much longer and analyze how the dissipation due to the perturbation in the hybrid region influences the actual numerical results.}
 In general, some systems, \blue{in the time window considered}, have shown a clear Hamiltonian nature, they are characterized by a relatively large HY region and thus the condition: $\frac{dw(x)}{dx} \ll 1$ is satisfactorily met and the CG region is large enough to provide thermodynamic stability to the rest of the system. Fig.\ref{temph2} show the temperature as a function of time compared with the equivalent full atomistic NVE simulation for system {\redc{(f)}}, taken as representative example; the agreement is remarkable (if the system was dissipative the temperature would show a sizeable drift \redc{compared to the reference full atomistic simulation}).
As discussed before, Fig.\ref{temph2} is not sufficient to justify the claim that the Hamiltonian of the system is a ``proper'' adaptive Hamiltonian. In fact, it may well be that while the system preserves the temperature in a proper way, other properties are instead modified in such a way that they do not reproduce the properties of the reference full atomistic system (or full coarse-grained system). For this reason we studied two important structural properties, that is the molecular number density and the radial distribution functions, C-C, C-H and H-H in the atomistic region; here we report the result for the most critical case that is for system $\blue{(6)}$.  Fig.\ref{rho1} shows for the system $\blue{(6)}$ the density across the box after $750 ps$. After such time if the system did not behave properly it should display large deviations from the reference NVE results. The density profile agrees with the reference one, with differences, in the most unfavorable case, of about $6\%$ in the region of maximal perturbation, that 
is in the HY region, however in the rest of the system the differences are below $2\%$ which is highly satisfactory. Moreover Fig.\ref{gofr} shows that the atomistic radial distribution functions of the adaptive systems essentially overlap with those of the reference full atomistic system. 
Other systems reported in Table \ref{tab2} and characterized as ``$H$'' show the same accuracy of system \blue{(6)}.
\label{beyond}

\begin{figure}[htbp]
\begin{center}
\includegraphics[width=6cm]{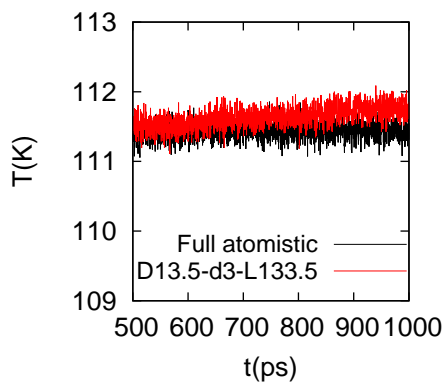}
\caption{Temperature as a function of time for system \blue{(2)}. Data are analyzed after 500 ps to allow the system for basic equilibration in the adaptive set up.} \label{H/D}
\end{center}
\end{figure}

\begin{figure}[htbp]
\begin{center}
\includegraphics[width=6cm]{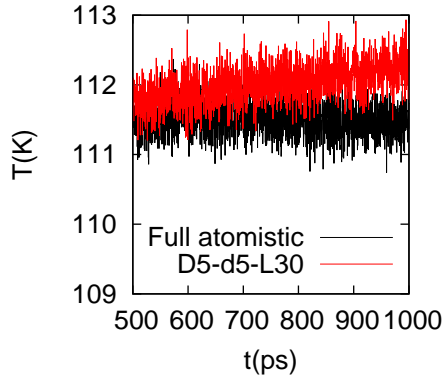}
\caption{Temperature as a function of time for system \blue{(7)}. Data are analyzed after 500 ps to allow the system for basic equilibration in the adaptive set up.} \label{D}
\end{center}
\end{figure}

On the basis of this result we can claim that we have found several systems where the adaptive set up corresponds closely to a self-contained Hamiltonian which produces highly satisfactory results on time windows of at least \blue{1000} ps.
\redc{As said before the time window considered is the same as that used in other adaptive work to calculate static properties of the full atomistic region.}
Instead systems {$\blue{(2)}$} and $\blue{(7)}$ do not produce satisfactory results; in system {$\blue{(2)}$} the transition region is very small and thus it strongly violates the mathematical condition of $w(x)$ being slowly varying. In fact despite the fact that the dissipation produced by the sharp transition can be mostly adsorbed by the large AT and very large CG region, the dissipation of energy is clearly visible in the plot of the temperature as a function of time (see Fig.\ref{H/D}). Finally for system $\blue{(7)}$ we have a smaller AT and CG region than for system {$\blue{(2)}$}; this, together with the small size of the HY region, makes the system highly dissipative as reported in Fig.\ref{D}. System $\blue{(8)}$ instead, although \redc{it} has got a relatively large HY region, is characterized by a CG region which is too small to act as equilibrating reservoir.
\blue{
\subsection{Hamiltonian behaviour v.s. dissipative behaviour: Beyond $1.0 ns$}
In the previous section we have chosen a rather strict criterion for defining a systems' behavior as $H$ or $NH$. However the strict criterion allows to state that in the time frame considered there are no differences between a full atomistic simulation and an adaptive resolution simulation. Nevertheless one needs to address the question of what happens beyond the time window considered, above all for systems classified as $H$; in fact the lost or gain of energy is cumulative, that is it adds up during the simulation. This process is inevitable since, beside the integration error present also in the full atomistic simulation, in the adaptive resolution simulation we have only an approximate Hamiltonian system. In this section we discuss this aspect by considering the three systems discussed in the previous sections. In particular for system (6), we have performed a much longer simulation ($4.0 ns$) and compared its results with the equivalent full atomistic simulation. The slope of the linear regression of the temperature v.s. time for systems (2) and (7) is respectively $6.6\time 10^{-4}$ and $1.15\time 10^{-3}$, that is the Kelvin dissipated (acquired) per $ps$. This implies that in a time window of $4000 ps (4.0 ns)$, system (2) increases the temperature of about $2.7 K$ while system (7) increases the temperature of $4.6 K$. The deviations found are not dramatic, but certainly sizeable. For system (6) instead, being classified as $H$ we have run a longer simulation and the curve of the temperature v.s. time is reported in Fig.\ref{long}.
\begin{figure}[htbp]
\begin{center}
\includegraphics[width=6cm]{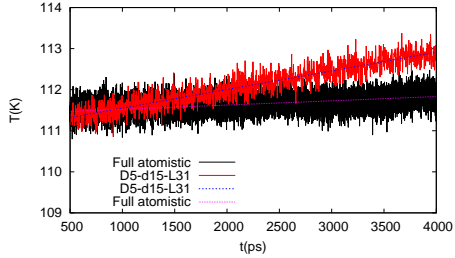}
\caption{Temperature as a function of time for system \blue{(6)} for $4.0 ns$. Data are analyzed after 500 ps to allow the system for basic equilibration in the adaptive set up. The continuous lines represent the curve of linear regression for, respectively, the adaptive resolution simulation and the corresponding full atomistic simulation.} \label{long}
\end{center}
\end{figure}
The slope in this case is $4.4 \time 10^{-4}$ and this should be also compared with the correspondent quantity of the full atomistic simulation which is $1.0 \times 10^{-4}$. The result is that the deviation from the ideal (target) temperature after $4.0 ns$ is of about $1.6 K$ while the difference with respect to a corresponding full atomistic simulation is about $1.3 K$. The question is now whether for (6) one may in practice use the longer simulation for calculating, e.g. structural properties. 
\begin{figure}[htbp]
\begin{center}
\includegraphics[width=6cm]{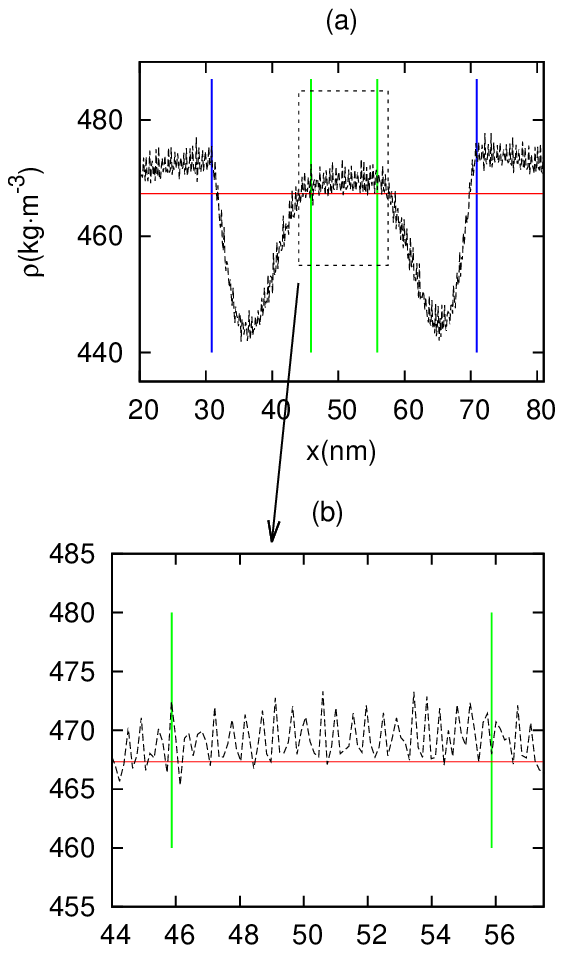}
\caption{\blue{(a)} Molecular number density as a function of the position in the simulation box of system \blue{(6)} in the time window $1.0-4.0 ns$. The horizontal line corresponds to the uniform target density. The blue vertical lines correspond to the borders between  the HY and CG region, the vertical green lines correspond instead to to the borders between  the HY and AT region. \blue{(b)} A zoom of the density in the AT region.} \label{denlong}
\end{center}
\end{figure}
\begin{figure}[htbp]
\begin{center}
\includegraphics[width=12cm]{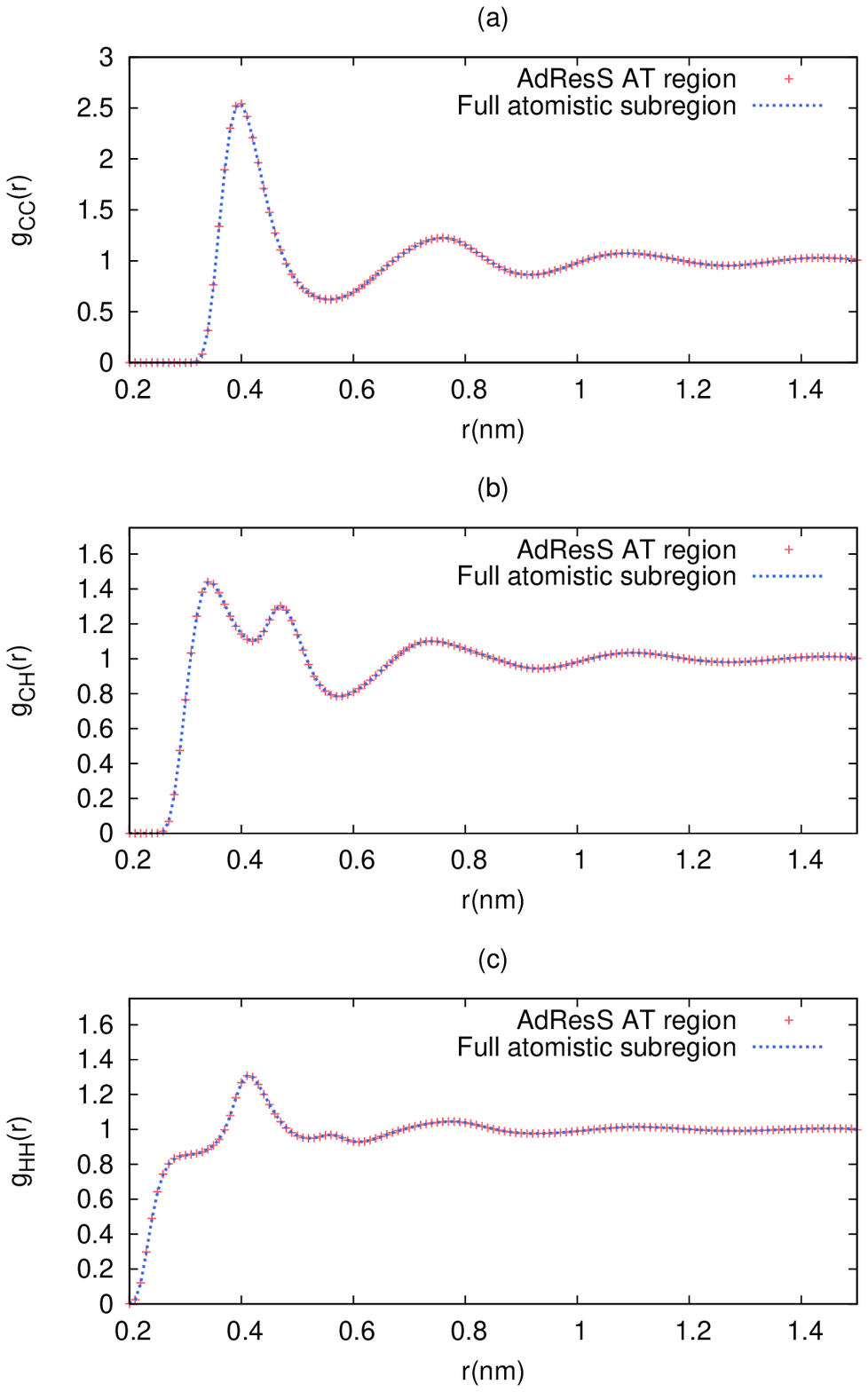}
\caption{Atom-atom radial distribution functions in the AT region compared with the equivalent from the full atomistic simulation of reference for system (6). Calculations are perform,ed over the time window $1.0-4.0 ns$.} \label{grlong}
\end{center}
\end{figure}
In Figs.\ref{denlong} and \ref{grlong} we have calculated the density across the system and various atomistic radial distribution functions in the atomistic region for a time frame between $1.0 ns$ and $4.0 ns$. This time frame considers time when the curve of the temperature of the adaptive resolution system and of the full atomistic system are not strictly overlapping anymore. The results show that the agreement of the structural properties is still satisfactory. For longer time windows of course system (6) will start to deviate significantly from a Hamiltonian behaviour. In any case, the results of this section show the robustness of approach for time windows which can be used for productive runs; certainly this approach represent a promising basis for building an improved numerical algorithm.} 
\blue{
\section{Discussion and Conclusion}
Hamiltonian-based approached to introduce the idea of adaptive molecular resolution are gaining popularity in the community of molecular dynamics. The conceptual background on which they are based is still subject of disputation; in particular the existence of a global adaptive resolution Hamiltonian written solely in terms of particles' degrees of freedom  is still an open question \cite{pre1,prlcomm,truh,entropy}.  In this work we have proposed a modification of the coupling between the atomistic and the coarse-grained region; the key difference with the other methods \cite{ensing1,hadress1} is that the Hamiltonian term which couples the atomistic region with the hybrid region and the Hamiltonian term which couples the coarse-grained region with the hybrid region are written in the form of full atomistic interactions and full coarse-grained interactions respectively without the introduction of any space dependent weight. This choice allows then to write the Hamiltonian corresponding to the interactions between the molecules of the hybrid region as a perturbation (introduced by a space-dependent interpolating weighting function) to the sum of a full atomistic and a full coarse-grained Hamiltonian. The perturbation then is a well defined term, independent of the interactions with the other regions, thus it can be mathematically analyzed in an asymptotic approach. From the technical point of view the partitioning proposed offers the possibility to implement the numerical procedure in a simplified and more general way for any molecular dynamics code: the atomistic and coarse-grained region can be treated independently and then synchronized by adding the perturbation term of the Hamiltonian.
We have carried out an asymptotic analysis and a numerical verification of the proposed scheme and show that when some mathematical conditions are reasonably fulfilled the system behaves practically as a 
Hamiltonian system. It should be underlined that the mathematical conditions are such that for molecular simulations in the current implementation the set up required by the Hamiltonian approach implies the choice of large systems and thus the necessity to perform large expensive calculations when compared to other adaptive Hamiltonian-like approaches \cite{ensing1,hadress1} or to  (technically) equivalent non-Hamiltonian/stochastic adaptive schemes based on the idea of open boundary/Grand Canonical-like approach \cite{jcpsimon,prlres,jctchan,prx,jcpanim1,njp,jcppi,pre2}. We hope that the conceptual and (potentially) technical advantages offered by our partitioning together with the detailed mathematical and numerical analysis carried here can be employed for improving the numerical and conceptual development of the adaptive resolution technique.}

\section*{Acknowledgment}
This research has been funded by Deutsche Forschungsgemeinschaft (DFG) through grants CRC 1114 (project C01) for RK and LDS and by the China Scholarship Council (file nr.201506010012) for JLZ.

\end{document}